\pdfoutput=1
\documentclass[conference,a4paper]{IEEEtran}

\usepackage{url}

\usepackage{textcomp}
\usepackage{import}
\usepackage{graphicx}
\usepackage{tikz}
\usepackage{amsfonts,amssymb,amsbsy}
\usepackage{gensymb}
\usepackage{amsmath}
\usepackage{amsthm}
\usepackage{multirow}
\theoremstyle{definition}
\newtheorem{definition}{Definition}[section]

\let\labelindent\relax
\usepackage{enumitem}
\usepackage{array}
\usepackage{caption}
\usepackage{subcaption}
\graphicspath{{./images/}}

\newcommand{\copyrighttext}{%
  \footnotesize \textcopyright 2016 IEEE. Personal use of this material is permitted.
  Permission from IEEE must be obtained for all other uses, in any current or future 
  media, including reprinting/republishing this material for advertising or promotional 
  purposes, creating new collective works, for resale or redistribution to servers or 
  lists, or reuse of any copyrighted component of this work in other works.}
\newcommand{\copyrightnotice}{%
\begin{tikzpicture}[remember picture,overlay]
\node[anchor=south,yshift=10pt] at (current page.south) {\fbox{\parbox{\dimexpr\textwidth-\fboxsep-\fboxrule\relax}{\copyrighttext}}};
\end{tikzpicture}%
}

\begin{document}

\title{Requirement verification in simulation-based automation testing}

\author{\IEEEauthorblockN{Eero Siivola$^1$\negthickspace, Seppo Sierla$^2$\negthickspace, Hannu Niemist\"o$^1$\negthickspace, Tommi Karhela$^{1,2}$ and Valeriy Vyatkin$^{2,3}$}
\IEEEauthorblockA{$^1$VTT Technical Research Centre of Finland Ltd\\
$^2$Dept. of Electrical Engineering and Automation, Aalto University, Espoo, Finland\\
$^3$SRT, Lule\r{a} Tekniska Universitet, Lule\r{a}, Sweden\\
 E-mail: \{eero.siivola, seppo.sierla\}@aalto.fi, \{hannu.niemisto, tommi.karhela\}@vtt.fi, vyatkin@ieee.org}}

\maketitle

\copyrightnotice
\IEEEpeerreviewmaketitle

\begin{abstract}
The emergence of the Industrial Internet results in an increasing number of complicated temporal interdependencies between automation systems and the processes to be controlled. There is a need for verification methods that scale better than formal verification methods and which are more exact than testing. Simulation-based runtime verification is proposed as such a method, and an application of Metric temporal logic is presented as a contribution. The practical scalability of the proposed approach is validated against a production process designed by an industrial partner, resulting in the discovery of requirement violations.
\end{abstract}
\section{Introduction \label{sec:introduction}}
The emergence of the Industrial Internet results in an increasing number of cyber-physical dependencies between the automation system and the process to be controlled; especially, the increasingly numerous and intelligent connected devices introduce complex temporal interdependencies that require advanced verification methods. There are multiple verification approaches, but traditionally three main techniques have been considered: theorem proving, model checking, and testing \cite{leucker2009brief}. Theorem proving, model checking and extensive testing do not scale very well to big systems. Runtime verification was developed to be a trade-off between model checking and testing to solve these problems \cite{leucker2009brief}. The research goal of this paper is to apply Metric temporal logic (MTL) to runtime verification of automation systems and to demonstrate it with an industrial case in mineral processing.

This paper is structured as follows. Section II introduces the basics of requirement formalization and runtime verification. Section III presents the theoretical approach. Section IV presents the case study, its implementation in a simulation environment and simulation results. Section V concludes the paper and identifies problems for further research.

\section{Literature review}
\subsection{Research goals}
Simulation-based testing is one active area for verification of industrial automation systems. In this paper, simulation-based runtime verification of real-time temporal logic requirements is supported, and the overall research goal in section \ref{sec:introduction} can now be elaborated to the following research goals:
\begin{itemize}
\item to be able to perform runtime verification of industrial automation systems before the physical system exists
\item to obtain scalability of the approach to an industrial-scale processes
\item to study the limitations of simulation based runtime verification based on an industrial scale case study
\end{itemize}

\subsection{Runtime verification}
\begin{figure}[t!]
  \centering
  \includegraphics[width=0.85\linewidth]{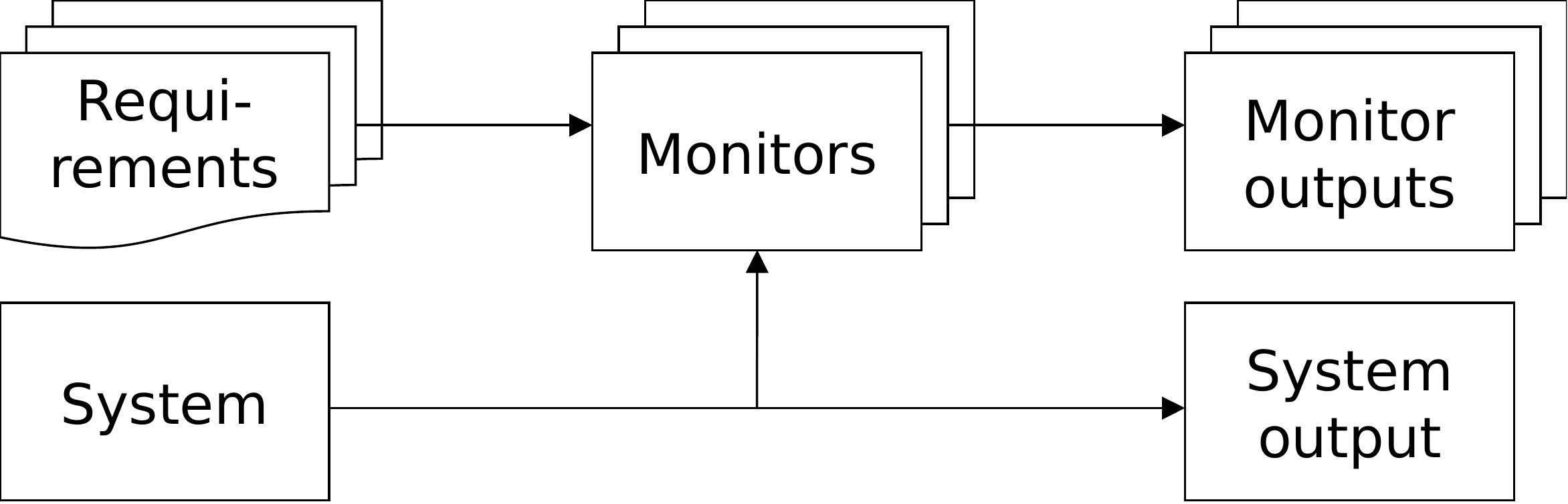}
\caption{Information flow in runtime verification.}
\label{fig:InformationFlow}
\end{figure}

Runtime verification is a lightweight and dynamic verification method used to determine whether a single finite run of a system violates given requirements \cite{colin200518}. The violation of requirements is checked with monitors, which are devices that read a set of states of a finite run and output a verdict \cite{levy2002combining}. A verdict is most often a value from a truth domain. The truth domain used in this work can be formulated as \{\textit{not violated}, \textit{violated}, \textit{not evaluated}\} \cite{schamai2013modelbased}. The information flow in runtime verification is shown in Fig \ref{fig:InformationFlow}.

 The monitors are automatically generated from the temporal logic requirements. One way for doing this is formula rewriting, where formalized requirements are updated at each time step during simulation. At each iteration every operator of the requirement is rewritten to correspond to the current state of the system by applying predefined set of rules. After this, the formulas are simplified to a canonical form using another set of rules \cite{havelund2001monitoring}\cite{havelund2004overview}.

\newcommand{\true}{\mathrm{true}}
\newcommand{\false}{\mathrm{false}}
\newcommand{\Z}{\mathbb{Z}}

\section{Theory}
\subsection{Metric Temporal Logic \label{sec:RuntimeVerification}}
The MTL formulation presented in this section mainly follows the formulation given by Thati \textit{et al.} in \cite{thati2005monitoring} with one major change explained at the end of this section. Given a finite set P of propositions $p$, that evaluate to either $\true$ or $\false$, MTL formulas are inductively defined as:
\theoremstyle{definition}
\begin{definition}{Let $\phi$ be a MTL formula that can be any of the following:}
\begin{equation*}
\phi := p \,|\, \wedge \{\phi_1,\ldots,\;\phi_n\} \,|\, \oplus \{\phi_1,\ldots,\;\phi_n\} \,|\, \circ_{I} \phi \,|\, \phi_1 \mathcal{U} _{I}\phi _2 
\end{equation*}
\label{def:MTL}
\end{definition}

\vspace*{-7mm}
Here $n \in \Z_{\geq 0}$ and names of the operators above are: 
$\wedge:$ and,
$\oplus:$ exclusive or,
$\circ_I:$ next, and
$\mathcal{U}_I:$ until.
Also, in above $p \in P$, and $I$ is an interval of non-negative real line.

A timed sequence $\rho$ is a sequence of pairs $(x_i, \tau_i)_{0\le i \le |\rho|}$ where $x_i$ is the state at $i^{th}$ time and $\tau_i$ is a non-decreasing sequence of times.
For given timed state sequence $\rho$ and index $1 \leq i \leq |\rho|$, let us define what it means for $(\rho,\;i)$ to satisfy the formula $\phi$. This can also be written as $(\rho,\;i) \models \phi$
\begin{align*}
 &(\rho,\;i) \models p & &\textnormal{iff } p \in x_i \\
 &(\rho,\;i) \models \wedge\{\phi _1 ,\ldots,\; \phi_n\}  & &\textnormal{iff } (\rho,\;i) \models \phi _j \textnormal{ holds for all} \\ & & &  1\leq j\leq n  \\
 &(\rho,\;i) \models \oplus \{\phi _1 ,\ldots,\; \phi_n\} & &\textnormal{iff } (\rho,\;i) \models \phi _j \textnormal{ holds odd number}\\&&&\textnormal{of times as $1\leq j\leq n$ }\\
 &(\rho,\;i) \models \circ _{I} \phi & &\textnormal{iff } i < |\rho|\textnormal{, }(\rho,\;i+1)\models \phi\\&&&\textnormal{and } t_{i+1} \in t_i +I \\
 &(\rho,\;i) \models \phi_1 \mathcal{U} _I \phi _2  & &\textnormal{iff } (\rho,\;j)\models \phi_2 \textnormal{ for some }\\&&& j \textnormal{ with } t_j \in t_i+I\textnormal{ and} \\
 & &  & (\rho,\;k)\models\phi_1\textnormal{ for all } i\leq k < j
\end{align*}
Operators $\wedge$ and $\oplus$ are defined as a set operators for convenience as this way logical values $true$ and $false$ follow straight from the definitions of $\wedge$ and $\oplus$. However, this notation makes equations harder to read. To make things simpler to follow, from this on let:
\begin{align*}
true& := \wedge \{ \} & false & := \oplus \{ \} \\ \phi &:= \wedge \{ \phi \} & \phi &:= \oplus \{ \phi \} \\
\end{align*}
\vspace*{-13mm}
\begin{align*}
\phi_1 \oplus ... \oplus \phi_n &:= \oplus \{\phi_1,\;...,\phi_n \}\\
\end{align*}
\vspace*{-13mm}
\begin{align*}
\phi_1 \wedge ... \wedge \phi_n &:= \wedge \{\phi_1,\;...,\phi_n \} \\
\end{align*}

\vspace{-5mm}
Using the primitive operators defined above, it is possible to define other common logic operators such as $\vee$: or, $\rightarrow$: implication, and $\neg$: not. Also, because temporal operators $\circ _I $ and $\mathcal{U}_I$ are not very intuitive, it often is useful to define additional temporal operators, the most used of which are $\lozenge_I$: finally and $\square_I$: globally. For these $(\rho,\;i)$ satisfies formula $\phi$:
\begin{align*}
&(\rho,\;i) \models \phi _1 \vee \ldots\vee \phi_n & &\textnormal{iff } (\rho,\;i) \models \phi _j \textnormal{ holds at least once} \\&&&\textnormal{as $ 1\leq j\leq n$ } \\
&(\rho,\;i) \models \phi _1 \rightarrow \phi_2 & &\textnormal{iff } (\rho,\;i) \models \phi _1\textnormal{ and }(\rho,\;i) \not \models \phi_2\\&&& \textnormal{do not both hold}\\
&(\rho,\;i) \models \neg \phi && \textnormal{iff }(\rho,\;i) \not \models \phi\\
& (\rho,\;i) \models \lozenge _I \phi & & \textnormal{iff } (\rho,\;i)\models \phi \textnormal{ for some } j \textnormal{ with}\\&&& t_j \in t_i+ I\textnormal{ and } i \leq j \leq |\rho|\\
& (\rho,\;i) \models \square _I  \phi & & \textnormal{iff } (\rho,\;i)\models \phi \textnormal{ for all } j \textnormal{ with}\\&&& t_j \in t_i+ I\textnormal{ and } i \leq j \leq |\rho| 
\end{align*}

\vspace{-3mm}
Which can be derived iteratively as follows:
\theoremstyle{definition}
\begin{definition}{Let $\phi$ be an additional MTL formula, these can be derived using already defined MTL formulas as:}
\begin{align*}
 \phi_1\vee\ldots\vee \phi_n =  & \neg \big( (\neg\phi_1)\wedge\ldots\wedge(\neg\phi_n)),\\
  \phi_1 \rightarrow \phi_2 = & \neg \big(\phi_1 \wedge ( \neg \phi_2) \big),\\
\end{align*}
\vspace*{-17mm}
\begin{align*}
 \neg\phi =  \true \oplus \phi, && \lozenge _I \phi = \true \; \mathcal{U}_I \phi, && \square _I \phi = \neg \big( \lozenge_I(\neg\phi)\big)\\
\end{align*}
\label{def:MTLadd}
\end{definition}
\vspace*{-11mm}

There exists alternative definition for MTL that substitutes $\oplus$ with $\neg$ and $\vee$. However, as one operator is replaced with two, the programmatic implementation becomes more comple. Furthermore, as MTL is presented this way, the mathematic formulation corresponds to the actual programmatic implementation used to obtain results in the case study. In addition to this, the used formulation of proportional logic is popular in the field of logic, where the origins of this work are.

From this on let $\circ := \circ_{[0,\;\infty)}$,  $\circ_{\leq t} := \circ_{[0,\;t)}$, $\circ_{>t} := \circ_{(t,\;\infty)}$ and similarly for other operators that have $I$. As a side note the most simple temporal logic, LTL can now be defined to be MTL, for which all $I=[0,\; \infty)$.

In addition to these operators defined above, as discussed in the literature review, it often is useful to introduce some additional operator templates for the most popular requirements. Kansas University's Laboratory for Specification, Analysis, and Transformation of Software (SAnToS) maintains a web repository \cite{kansas2015web} for commonly used and well tested LTL patterns. We noticed that these can easily be extended to MTL. The most useful pattern in the context of our research was noticed to be: 
\begin{align*}
     &\mathrm{timedTrigger}_{I}(\phi_1,\;\phi_2)= \\& 
     \square  
     \Big(
         \big( ( \neg \phi_1 ) \land ( \circ \phi_1 ) ) \rightarrow
         \circ (\lozenge_{I} \phi_2 )
      \Big)
\end{align*}

\vspace{-3mm}
The above template can be used to formalize two useful requirement. If $I=[0,\; \infty)$, the template corresponds to requirement "\textit{When $\phi_1$ becomes true, $\phi_2$ must eventually  become true}". If $I=(t_1,\;t_2]$, the template corresponds to requirement "\textit{When $\phi_1$ becomes true, $\phi_2$ must become true within $t_1$ to $t_2$ seconds}"

Unlike in the formulation given by Thati \textit{et al.} in \cite{thati2005monitoring} and original formulation of MTL by Alur \textit{et al.} in \cite{alur1990real}, past time temporal operators that correspond to operators $\circ$ and $\mathcal{U}$ are not considered in our research to be part of MTL. Although it is known that introducing these two past operators makes MTL formulas more expressive, no evident advantage from introducing them would had been gained at least for the requirements formalized in the test process \cite{bouyer2005on}. Similar results from LTL, where only future time operators have been sufficient for presenting requirements can be found from \cite{tommila2014controlled} and \cite{dwyer1999patterns}.

\subsection{Term rewriting based runtime monitoring for MTL}

Let $\phi\{\rho,\;i\}$ symbol the derivation of MTL formula $\phi$ with respect to the $i^{th}$ event of timed state sequence $\rho$. This derivation process for different MTL formulas is iteratively defined as follows:
\theoremstyle{definition}
\begin{definition}{Let $\rho$ be timed state sequence and \\ $1 \leq i \leq |\rho|$, $i\in \mathbb{N}$. Then:}
\begin{align*}
 &p  \{\rho,\;i\} =  p \in \pi _i\\
  (\phi_1\wedge\ldots\wedge\phi_n&)  \{\rho,\;i\} =  \phi_1 \{\rho,\;i\}\wedge\ldots \wedge \phi_n \{\rho,\;i\} \\
 (\phi_1\oplus\ldots\oplus\phi_n&)  \{\rho,\;i\} =  \phi_1 \{\rho,\;i\}\oplus\ldots\oplus \phi_n\{\rho,\;i\} \\
 (\circ _ {I} \phi &) \{\rho,\;i\} = \\  &\phi_1 \left\{ \begin{array}{ll}
         \phi & \mbox{if $\tau_{i+1} \in \tau_i + I$ and $i < |\rho|$} \\
        false & \mbox{else}.\end{array} \right.\\
 ( \phi_1 \mathcal{U}_I \phi_2&) \{\rho,\;i\}= \left( \left\{ \begin{array}{ll}
         \phi_2\{\rho,\;i\} & \mbox{if $0 \in I$} \\
        false & \mbox{else}\end{array}\right. \right) \\
        & \vee \left(
       \left\{ \begin{array}{ll}
          (\phi_1\{\rho,\,i\} ) \wedge ( \phi_1 \mathcal{U}_{I'}\phi_2) & \mbox{if $i < |\rho|$}\\
        \false & \mbox{else}
        \end{array} \right.           
        \right)\\
        &\mbox{here $I' = I-\tau_{i+1}+\tau_{i}$} 
\end{align*} \label{def:monitor1}
\end{definition}

\vspace*{-5mm}
In order to make the algorithm more efficient, canonical term-rewriting system for Boolean algebra developed by Hsiang in \cite{hsiang1985refutational} is used.
This algorithm keeps the formula in the algebraic normal form where equivalent formulas have the same representation.
The algorithm is defined as a term-rewriting system defined above. Because subformulas of the connectives are represented as sets, associativity and combination of subformulas with the same representation is automatically taken care of.

\begin{definition}{Let $\phi$ be any MTL formula. Then the equations on the left side of $\Rightarrow$ are transformed to the equations on its right side:}
\begin{align*}
&\phi \wedge \false \Rightarrow \false  && \phi \wedge \phi \Rightarrow \phi && \phi \wedge \true \Rightarrow \phi  \\
&\phi \oplus \false \Rightarrow \phi && \phi \oplus \phi \Rightarrow false &&
\end{align*}
\vspace{-\belowdisplayskip}
\vspace{-\abovedisplayskip}
\vspace{-3\jot}
\begin{align*}
\phi_1 \wedge (\phi_2 \oplus \phi_3) \Rightarrow (\phi_1 \wedge \phi_2) \oplus (\phi_1 \wedge \phi_3)
\end{align*}
\label{def:MTLtransformation}
\end{definition}


%
\vspace*{-6mm} 

The algorithm presented in this section mainly follows the one presented by Thati \textit{et al.} in \cite{thati2005monitoring} with two exceptions. Firstly, the past operators are not included in this algorithm because, as explained in the previous section, no evident advantage from using them would be gained. Secondly, the algorithm by Thati \textit{et al.} transforms the MTL formulas to binary presentations and performs the evaluation and canonisation in this form. Binary presentation would make the algorithm more efficient if also the past form of MTL was used. 
\section{Case study}
\subsection{Pressure leaching process}
The tool based on the above algorithm is tested on a simulation model of a real industrial mineral processing plant. The whole process, where copper, cobalt and zinc are extracted from thermal treated limestone, known as calcine, is far too complex to be used as a test process. Therefore only a subprocess of a subprocess called pressure leaching is used. The whole process is described briefly in the next paragraph in order for the reader to find the context of the chosen system. After this, one subprocess, autoclave, of pressure leaching is described in more detailed manner.

Calcine is first crushed, grinded and then mixed with water in the subprocess called pulping. The resulting solution called slurry is then pumped to the pressure leaching process where metals are dissolved into the solution and pure lime is extracted from it. The leach residue, pregnant leach solution (PLS), that has high concentration of metals is then processed in three subsequent phases in each of which one metal is extracted from the slurry. After this, all compounds are end processed in suitable ways resulting copper cathodes, cobalt carbonate, zinc carbonate and neutralised waste.
\begin{figure}[t]
  \centering
  \includegraphics[angle=270, width=0.95\linewidth]{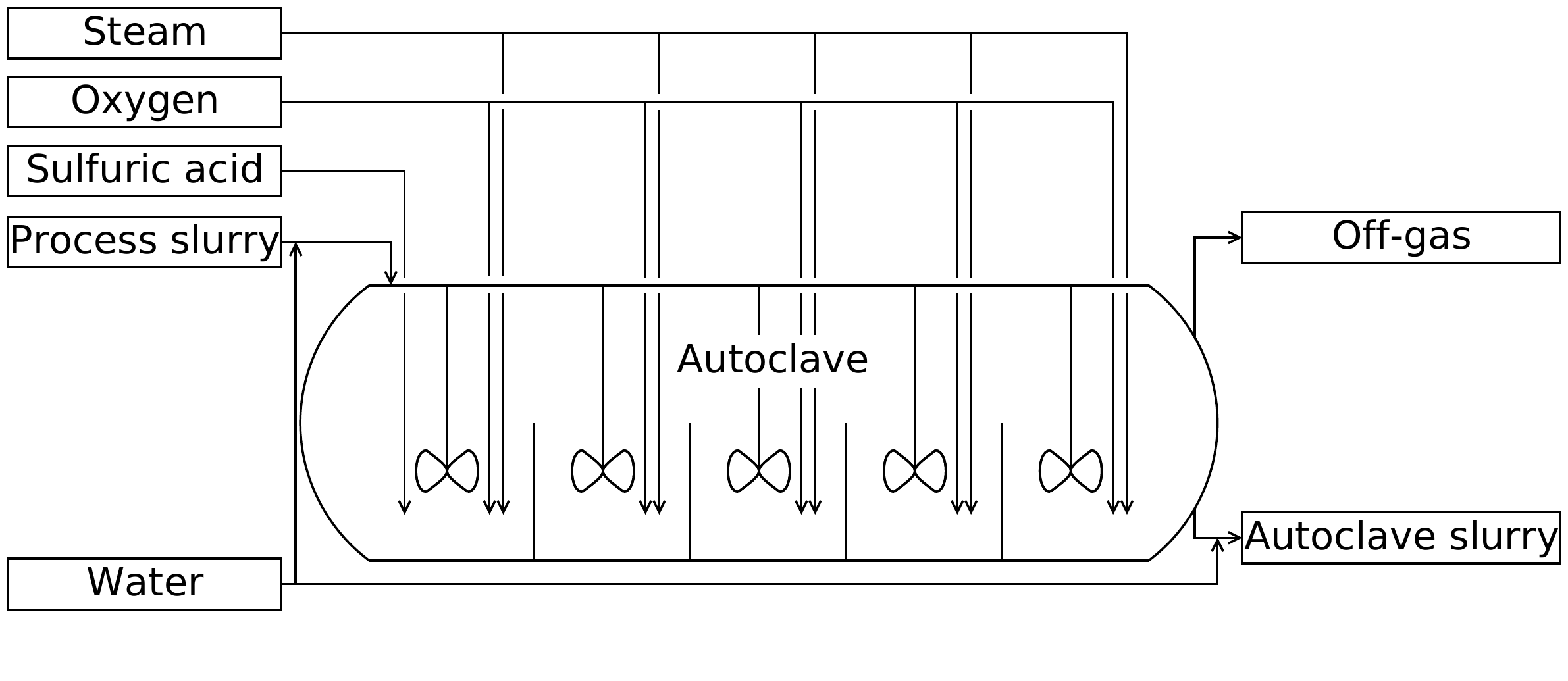}
\caption{Subprocess of pressure leaching: metals dissolve to process solution inside an autoclave.}
\label{fig:autoclave}
\end{figure}

In considered subprocess, autoclave, the chemical reactions happen and metals dissolve to heated calcine slurry. This process slurry is fed to an autoclave that consists of five compartments each of which is equipped with oxygen input, steam input and agitator. The oxygen partial pressure is kept at 5 bar to maintain the oxidative conditions for reactions. The absolute pressure in autoclave is 28 bar. Steam is injected to maintain the temperature of 220 $^\circ$C. In addition to these injections, sulphuric acid concentration of 40--50~g/L is maintained inside the autoclave by injecting it to the first compartment to keep the solution acidic enough for metals to dissolve. Simplified flow diagram of the autoclave is illustrated in Fig \ref{fig:autoclave}.

The process and its automation were modelled using a tool called Apros, which is built using software implementation platform called Simantics \cite{karhela2012open}.

\subsection{Elicited requirements}
26 formalized requirements imposed on the automation of the process of the case study can be divided to the following two groups:
\begin{enumerate}
\item Limit requirements: This group contains some absolute limits that the process variables must never exceed. E.g. "\textit{Pressure of all Autoclave compartments must always be between absolute pressure of 27.9 and 28.1 bar}". 20 of this kind of requirements were formalized. These requirements are related to the upper and lower limits of slurry, temperatures and pressures, and to mass flows in some pipes. These requirements can be formalized as:
\begin{equation*}
\square p
\end{equation*}
Here p is a function that returns true if the wanted state is within bounds.

\item Order related requirements: This group contains requirements for order of two events. These requirements can further be divided into two categories:
\begin{enumerate}
\item Latter event must happen within some time limit. E.g "\textit{If the pressure difference between steam feed line and autoclave is less than 0.1 bar, the steam feed line valve must close within 60 seconds}". 5 of this kind of requirements were formalized. These requirements can be formalized as:
\begin{equation*}
\mathrm{timedTrigger}_{[0,\;t]}(p_1,\;p_2)
\end{equation*}
Here t is the time limit, $p_1$ is a function returning true when the first event happens and $p_2$ is a function that returns true when the latter event happens.

\item The latter event must happen after a delay. E.g. "\textit{If the slurry level in Autoclave goes under the limit of 1 meter, then the output valve must not close before 30 seconds, but must be closed after 60 seconds}". One of this kind of requirement was formalized. These requirements can be formalized as:
\begin{align*}
& \big (\neg \mathrm{timedTrigger}_{[0,\;t_1]}(p_1,\;p_2)\big) \\ & \wedge  \mathrm{timedTrigger}_{[t_1,\;t_2]}(p_1,\;p_2)
\end{align*}
Here $t_1$ is the time limit before which the latter event should not happen, $t_2$ is the time limit before which it should happen, $p_1$ is a function returning true when the first event happens and $p_2$ is a function returning true when the latter event happens.
\end{enumerate}

\end{enumerate}

\subsection{Results}
The selected tests and requirement verification results are presented in TABLE \ref{tab:tests}. Most notably, autoclave control seems to be sensitive to malfunction of off-gas valve. Instantly when the malfunction occurs, the temperature inside all autoclave compartments starts to drop as the steam input valves are closed. Simultaneously partial pressure of the oxygen becomes too large and 7 requirements are violated. 

\begin{table}[t!]
\begin{center}
\caption{Description of performed tests, what they are testing and number of requirements violations. \label{tab:tests}}
\setlength\tabcolsep{2mm}.
\begin{tabular}{ p{6.6cm}  p{1cm}}
\hline
Test description & Violations\\[-0.5mm]
 \hline \\[-7pt]
 Simulation ran normally & 0 \\[5pt]
Simulation model started & 1 \\[5pt]
Deviation in the pressure measurement of the first compartment  & 0\\[14pt]
Malfunction in the oxygen feed valve of the first compartment  & 0\\[14pt]
Autoclave temperature measurement failure in the second compartment  & 1\\[14pt]
Large pressure drop in the steam feed valve of the third compartment  & 0 \\[14pt] 
Malfunction in the autoclave off-gas valve   & 7\\[5pt]
Bias added to the level measurement in the last compartment & 2 \\
   \hline
\end{tabular}
\end{center}
\end{table}

\section{Conclusions and Further Work}

The requirements verification method could reveal errors in the automation of the process even though the test case selection was not based on process knowledge but rather on typical use cases and some of the most probable error sources. Also, it could reveal requirement violations caused by interconnections within subprocess, which are hard to find with traditional automation testing methods. 
The implemented requirement monitors did not slow down the simulation even when all 26 monitors were run simultaneously. No requirements impossible to formalize with MTL were found in this case study. In addition to this, most of the requirements could be grouped similarly as found by Dwyer \textit{et al.} in \cite{dwyer1999patterns}.

\section*{Acknowledgment}
This research has been done as a part of S-STEP project, which is funded by Finnish Metals and Engineering Competence Cluster (FIMECC). Furthermore, the authors would like to thank Juha Kortelainen and Tuomas Miettinen from VTT Technical Research Centre of Finland Ltd.

\bibliographystyle{IEEEtran}
\bibliography{indinbib}

\end{document}